\documentclass[conference]{IEEEtran}
\IEEEoverridecommandlockouts

\usepackage{graphicx}
\usepackage{subcaption}
\usepackage{cite}
\usepackage{amsmath,amssymb,amsfonts}
\usepackage{cleveref}
\usepackage{algorithmic}
\usepackage{graphicx}
\usepackage{textcomp}
\usepackage{xcolor}
\usepackage{todonotes}
\usepackage{xcolor}
\definecolor{darkgreen}{rgb}{0,0.4,0} 
\definecolor{softyellow}{RGB}{255, 204, 102}
\usepackage{pifont}
\usepackage{multirow}
\usepackage{tabularx}
\usepackage{url}
\usepackage{array}
\usepackage{footmisc}
\usepackage[most]{tcolorbox}
\usepackage{adjustbox} 
\usepackage{makecell}
\usepackage{multirow}
\usepackage[table]{xcolor}
\definecolor{lightyellow}{RGB}{255,250,205}

\AtBeginDocument{%
  \providecommand\BibTeX{{%
    Bib\TeX}}}

\def\BibTeX{{\rm B\kern-.05em{\sc i\kern-.025em b}\kern-.08em
    T\kern-.1667em\lower.7ex\hbox{E}\kern-.125emX}}

\usepackage[most]{tcolorbox}
\tcbuselibrary{skins, breakable}

\newtcolorbox{lessonbox}[2][]{%
  enhanced,
  colback=blue!5,
  colframe=blue!60!black,
  coltitle=black,
  colbacktitle=white,
  fonttitle=\bfseries,
  title=#2,
  sharp corners,
  boxrule=0.6pt,          
  width=\linewidth,
  breakable,
  boxsep=2pt,             
  left=4pt, right=4pt,    
  top=2pt, bottom=2pt,    
  before skip=6pt,        
  after skip=6pt,         
  #1
}

\usepackage{tcolorbox}
\usepackage{enumitem}

\newtcolorbox{lessonlist}{
    colback=blue!5!white,
    colframe=blue!40!black,
    boxrule=0.5pt,
    arc=2pt,
    left=4pt,
    right=4pt,
    top=4pt,
    bottom=4pt
}

\newcommand{\polaris}{Polaris}
\newcommand{\aurora}{Aurora}

\begin{document}

\title{When More Cores Hurts: The Vector Database Scaling Paradox in HPC}

\author{
Seth Ockerman$^{1}$,
Song Young Oh$^{2}$,
Amal Gueroudji$^{3}$,
Rochana Chaturvedi$^{3}$,
Philip Carns$^{3}$, \\
Nicholas Chia$^{3}$,
Matthieu Dorier$^{3}$,
Robert Latham$^{3}$
Tanwi Mallick$^{3}$,
Swan Perarnau$^{3}$,\\
Robert Underwood$^{3}$,
Kyle Chard$^{2,3}$,
Ian Foster$^{3,2}$,
Robert Ross$^{3}$,
Shivaram Venkataraman$^{1}$\\[0.5em]
$^{1}$University of Wisconsin--Madison \quad
$^{2}$University of Chicago \quad
$^{3}$Argonne National Laboratory
}

\maketitle

\begin{abstract}

Vector databases have been designed and optimized for cloud environments; however, emerging scientific AI workloads (e.g., molecular search, meteorological trajectory detection, and literature-driven hypothesis generation) demand efficient, scalable execution on HPC systems. We present a large-scale evaluation of three state-of-the-art vector databases---Qdrant, Milvus, and Weaviate---on two production supercomputers, scaling to 256 distributed workers across 64 compute nodes. We evaluate representative workload patterns---mixed read/write and write-then-read---using popular benchmarks, multimodal embeddings, and a novel real-world scientific dataset. Our results reveal that workload characteristics can limit latency reduction, additional cores can reduce query throughput by up to 30.67\%, and scaling from 16 to 256 workers (16$\times$) only yields a 5.46$\times$ improvement. This scaling paradox exposes the fundamental mismatch between cloud-oriented designs and HPC systems, highlighting the need for new, HPC-aware vector database designs.

\end{abstract}





\section{Introduction}
Vector databases (VDBs) have emerged as a crucial tool for managing and querying unstructured data. VDBs store compact vector representations of data known as embeddings~\cite{mikolov2013efficientestimationwordrepresentations, pennington-etal-2014-glove} and facilitate efficient semantic search, powering applications such as recommendation systems~\cite{Hu2026_Gen,Akhiiezer2025_Curriculum}, 
retrieval-augmented generation~\cite{Smurf2025, sarmah2024hybridragintegratingknowledgegraphs, fan2024surveyragmeetingllms}, and long-term agentic memory~\cite{packer2024memgptllmsoperatingsystems,jiang2026magmamultigraphbasedagentic,zhong2023memorybankenhancinglargelanguage,xu2025amem}. These traditional AI workloads, as well as emerging scientific use cases such as cosmology~\cite{xia2025multimodalfoundationmodelcosmological}, earth science~\cite{coulaud2026traknnefficienttrajectoryaware}, and medicine/biology~\cite{Yang2025Dual,ockerman2025exploringdistributedvectordatabases,lu2023foundationalmultimodalvisionlanguage,Ji2024Automating}, require scalable VDB performance on high-performance-computing (HPC) systems. Scientific datasets are increasingly large, already reaching petabyte scale \cite{Erying2016, Fairley2019, cappelloWhatSupportWhen2025}; and, concurrently, researchers are investigating representing these datasets as embeddings for semantic discovery and AI use~\cite{Yang_Learned2018, Ko_Scalable2026,genes15010025}. These developments underscore the need for the HPC community to understand what is required to operate VDBs at HPC scale.

VDB development has been driven primarily by industry~\cite{Wang2021Milvus, vespa2025, qdrant2025, vald2025, weaviate2025, johnson2017billionscalesimilaritysearchgpus}, with a focus on cloud-based deployments~\cite{li2025cloudnativevectorsearchcomprehensive}. 
As a result, 
to the best of our knowledge, little prior work has investigated, or optimized, performance of VDBs on HPC systems. HPC systems differ significantly from typical cloud platforms in terms of performance~\cite{Lange2023_CloudHPC, sochat2025usabilityevaluationcloudhpc, Munhoz2023APerformance} and architecture~\cite{Latham2025Initial, Kwack2025AIandHPC, HomerdingPolarisAA}, featuring many-core CPUs, multiple high-memory GPUs per node, a tightly coupled architecture connected by a high-bandwidth network, and a deep storage hierarchy~\cite{Liang2020DAOS}. This gap motivates our study, which evaluates the behavior of state-of-the-art VDBs on HPC systems and examines whether their designs align with the requirements of real-world scientific workloads.

We evaluate three popular VDBs---Weaviate~\cite{weaviate2025}, Milvus~\cite{Wang2021Milvus}, and Qdrant~\cite{qdrant2025}---using four embedding/query datasets (Pes2o-VE, Yandex-text-to-image~\cite{Yandex2016_Deep1B}, GIST~\cite{Jegou2011Product}, and dbpedia-openai-1M~\cite{dbpedia-entities-openai-1M}) on two production supercomputers. To support our evaluation and promote reproducibility, we release \texttt{VECHINI},\footnote{\url{https://github.com/OckermanSethGVSU/VECHINI}} a configurable framework for benchmarking VDBs on HPC platforms. Additionally, we construct and release Pes2o-VE,\footnote{\url{https://www.materialsdatafacility.org/detail/e31e3225-7f75-4313-9983-f8b75811405f-1.0}} a real-world embedding dataset inspired by a biological application, which, to the best of our knowledge, is among the largest publicly available embedding datasets ($\approx$88 million embeddings or 843.56~GB). Using VECHINI, we study two common VDB workload patterns: (i) insert-then-query, representative of scientific and knowledge-driven workflows such as hypothesis generation and literature retrieval, and (ii) mixed insert-query, characteristic of emerging AI and agentic applications~\cite{yan2025mofadiscoveringmaterialscarbon, hellert2025ospreyproductionreadyagenticai}. Our study provides the following valuable insights into the performance of current VDBs on HPC systems: 

\textbf{Cloud vs HPC:}  We demonstrate that, compared with the cloud, query throughput and index time improve the most on HPC systems, while already low-latency queries see minimal improvement. 

\textbf{Impact of Embedding geometry:} Complex embedding spaces that require deeper search to meet recall targets~\cite{aumuller2019rolelocalintrinsicdimensionality, Houle2018On, Elliott_2024} benefit more from HPC architectures, achieving greater reductions in latency and improvements in throughput. 



\textbf{ Limits of the Streaming-based Insertion Model:} VDB architectures fail to exploit the hierarchical storage landscape of HPC (e.g., local SSD, DAOS, Lustre) and enforce fine-grained consistency even during bulk ingestion, leaving substantial resources unused.

\textbf{Hidden Costs at Scale:} VDB systems incur write-amplification, leading to significant storage overhead (2.8$\times$ with Milvus) that may hinder scalability for large datasets. 

\textbf{Impact of Segmentation:} Traditionally, recall–cost trade-offs are tuned through index parameters; in modern VDBs, however, the system’s data partitioning strategy also acts as an implicit tuning axis, jointly shaping recall, throughput, and latency (e.g., 1 vs. 8 partitions improved recall from 0.962 to 0.985 but increased search time from 4.19s  to 22.65s.)

\textbf{Limited or Inverse Scaling:} Across both single-node and distributed settings, adding cores or nodes yields diminishing returns in query throughput and, in some cases, even degrades query performance by up to 30.67\%.


In summary, we study multiple state-of-the-art VDBs with up to 256 distributed VDB workers across 64 compute nodes, exploring both intra- and internode parallelism. We examine how VDBs interact with key HPC architectural features, including many-core CPUs, GPU acceleration, and DAOS/Lustre storage~\cite{Latham2025Initial,braam2019lustrestoragearchitecture}. Using our experience, we provide practical insights that aid scientists in using existing VDBs on HPC systems and highlight key areas for new research. 

\section{Background}

VDBs are specialized databases designed for efficient search over vectorized data representations known as embeddings, where embeddings encode semantic features about the data~\cite{qwen3embedding, lee2025nvembedimprovedtechniquestraining}. To perform a search, VDBs compute distances between a query vector and stored embeddings to identify the top-k nearest candidates, a process known as nearest-neighbor search~\cite{Muja2014ScalableNN,Gionis1999Similarity,Yianilos1993}. As the number of embeddings grows, exhaustive comparison becomes infeasible; instead, VDBs use approximate-nearest-neighbor (ANN) search~\cite{Indyk1998_ANN} with index structures~\cite{Bentley1975Multidimensional,Gutman1984,Mackenzie2025_Efficent,Andoni2008_LSH,jafari2021surveylocalitysensitivehashing} that provide controllable accuracy–latency trade-offs. Accuracy is measured primarily  by \texttt{recall@k} (see \Cref{eq:recall_at_k}), defined as the fraction of the $k$ true nearest neighbors returned by a top-k ANN query.


\subsection{Graph-Based Indices}
Graph-based indices\cite{subramanya2019diskann,fu2025fastapproximatenearestneighbor,ootomo2024cagrahighlyparallelgraph}, especially hierarchical navigable small world (HNSW) graphs~\cite{malkov2018efficientrobustapproximatenearest}, are the most popular choices for VDB indexing. HNSW graphs provide strong performance and the ability to preserve neighborhood structure in high-dimensional spaces. To control the trade-off between latency and accuracy, HNSW uses three key parameters: $M$, $ef_{\text{construction}}$ (EF-C), and $ef_{\text{Search}}$ (EF-S). $M$ controls the maximum node connectivity, while EF-C and EF-S refer to the size of the candidate queue used during index construction and search, respectively. The latter two parameters act as a proxy for index/search effort, with higher values indicating a more exhaustive process.


\begin{equation}
\mathrm{Recall@}k
= \frac{\left| Results_k
\cap GroundTruth_k \right|}{k}
\label{eq:recall_at_k}
\end{equation}


\subsection{Distributed Vector Databases }
\label{sec:dvb}
While HNSW and more recent GPU-based graph indices~\cite{ootomo2024cagrahighlyparallelgraph,rapidsai} provide strong performance on a single node, their effectiveness is ultimately bounded by node-level resource constraints. HPC workloads, by contrast, operate at scales that exceed the capabilities of any single node. To reflect the requirements of HPC environments, we restrict our evaluation to VDBs that support distributed computation.




While Milvus, Weaviate, and Qdrant differ architecturally, they share several core design patterns: a broadcast–gather query model, consistent hashing for write routing~\cite{Stoica2001Chord}, and segmented data layouts. In order to support queries over distributed indices, queries are broadcast to all relevant workers, which execute ANN search locally The local results are then aggregated into a global result. For new data, consistent hashing routes inserts to shards and appends the data to in-memory open segments. Once a segment reaches a predefined size threshold, it is marked as eligible for indexing. At this point, Milvus flushes the segment to storage for later indexing, whereas Qdrant and Weaviate build the segment index directly in memory. This distinction reflects a broader architectural divide. Qdrant and Weaviate use a worker-centric design in which each node manages the full life cycle of its data (insertion, indexing, query). In contrast, Milvus decomposes responsibilities across microservices and explicitly separates compute/storage. The front-ends of Milvus are proxies, which route data to the correct component and perform the final level of query aggregation. To process incoming data, Milvus uses streaming nodes, which maintain open segments in memory, perform exhaustive search over unindexed data as needed, and flush segments to durable storage (typically MiniIO or remote S3) when they reach a size threshold. For indexing and querying, Milvus utilizes data and query nodes, respectively, with both loading the relevant segments from durable storage into their local memory.

\begin{table*}[t]
\caption{Datasets used in this study listed from smallest to largest. All vectors are stored using \texttt{float32} representations.}
\centering
\label{tab:datasets}
\begin{tabular}{|c|c|c|c|c|c|c|}
\hline
\textbf{Dataset} & \textbf{\# Embeddings} & \textbf{Dim} & \textbf{Total Size (GB)} & \textbf{Use Case} & \textbf{Model(s)} & \textbf{Modality} \\
\hline
GIST   & 1,000,000 & 960   & 3.58~GB   & Image Search & N/A & Image \\
\hline
dbpedia-openai-1M & 1,000,000 & 1536 & 5.72~GB &  Information Retrieval & Text-embedding-ada-002 & Text \\
\hline
Yandex-text-to-image  & 1,000,000,000 & 200  & 745.05~GB & Multimodal Information Retrieval & Se-ResNext-101, DSSM & Image, Text \\
\hline
Pes2o-VE &  88,453,763 & 2560  & 843.56~GB & Academic Corpus Review & Qwen3-Embedding-4B & Text \\
\hline
\end{tabular}
\end{table*}

\subsection{Related Work} 
Vector databases are an active area of research, with multiple survey and evaluation efforts. Several recent works~\cite{Vector_Cuzzorcrea2025, taipalus2024vector,pan2023surveyvectordatabasemanagement,han2023comprehensive} summarize key challenges, design choices, and opportunities in vector database systems; however, they do not provide empirical evaluations. Most empirical evaluations, across both academia ~\cite{shen2024understandingsystemstradeoffsretrievalaugmented,Wang2021graphSurvey,aumuller2018annbenchmarksbenchmarkingtoolapproximate,Pandit2025_EvalANN,li2026ragperfendtoendbenchmarkingframework} and industry\cite{weaviate_ann_benchmark,qdrant_benchmarks,milvus_vdbbench_2025}, focus on centralized, single-node VDB systems with a single index rather than on distributed systems. While  a growing body of work is focused on improving distributed VDBs~\cite{xu2025scalabledistributedvectorsearch,zhi2025efficientscalabledistributedvector,xu2025harmony,Hu_2025}, these studies evaluate query-time performance in isolation, rather than the full workload life cycle---spanning insertion, indexing, and querying---or the properties of the distributed cloud systems on which they are deployed. In scientific workflows such as large-scale simulations \cite{Heitmann_2019,wang2025kilometerscalee3smlandmodel}, data is continually generated and can reach petabyte scale. In this setting, insertion and indexing performance are critical to determine whether  the system can keep pace with data generation, highlighting the need to move beyond query-time performance alone.

Recent work~\cite{li2025cloudnativevectorsearchcomprehensive} examines the performance characteristics of two index types in a cloud environment. The results highlight the impact of cloud storage and limited network throughput on search performance and index design. This work aims to uncover similar insights; however, we focus on HPC environments and evaluate three distributed vector database systems end-to-end, rather than comparing individual index structures. Recent work~\cite{ockerman2025exploringdistributedvectordatabases} evaluates Qdrant on an HPC system. However, it considers only a single system and VDB, does not exercise advanced features such as GPU acceleration, and does not explore HPC-specific architectural components (e.g., DAOS or Lustre). This work provides a more systematic evaluation across multiple HPC systems, distributed VDB implementations, and datasets, including Pes2o-VE, which is significantly larger than datasets used in prior VDB workload studies \cite{ockerman2025exploringdistributedvectordatabases,li2025cloudnativevectorsearchcomprehensive,li2026ragperfendtoendbenchmarkingframework}.





\begin{table}[t]
\centering
\caption{Overview of \polaris{} \& \aurora{} compute nodes.}
\label{tab:systems}
\begin{tabular}{|c|c|c|}
\hline
\textbf{Component} & \textbf{\polaris{}} & \textbf{\aurora{}} \\
\hline
CPU & 32 cores & 104 cores \\
\hline
Memory & 512~GB DDR4 & 1~TB DDR5 \\
\hline
GPUs & 4$\times$40 GB NVIDIA A100 & 6$\times$128 GB Intel Max \\
\hline
Node-local storage & 512~GB SSD & N/A \\
\hline
Network & 2$\times$25~GB/s NIC & 8$\times$25~GB/s NIC \\
\hline
Networked Storage & Lustre PFS & Lustre PFS \& DAOS \\
\hline
\end{tabular}
\vspace{-5pt}
\end{table}

\section{Methodology}
\label{sec:methodology}
We conduct our evaluation on two HPC systems: \polaris{},\footnote{\url{https://www.alcf.anl.gov/polaris}} which is equipped with NVIDIA A100 GPUs, and \aurora{},\footnote{\url{https://www.alcf.anl.gov/aurora}} which is equipped with Intel PVC GPUs. For scaling experiments, we default to \aurora{}, which offers a newer system architecture and greater overall computational capacity. GPU-based evaluations are performed exclusively on \polaris{} because of limited support for Intel GPUs in the evaluated VDBs. An architectural overview of both HPC systems is provided in \Cref{tab:systems}. Cloud deployments of VDBs typically rely on Docker and Kubernetes, which are not supported on most HPC systems, including \polaris{} and \aurora{}. Instead, we use Apptainer,\footnote{\url{https://apptainer.org/}} which converts Docker images into an HPC-compatible format, with Qdrant 1.16.2, Milvus 2.6.6, and Weaviate 1.36.0.

Milvus is designed as a cloud-native distributed VDB that assumes shared persistent state via object storage (e.g., S3/MinIO) or local disk in single-node deployments. However, based on recommendations from Milvus developers,\footnote{\url{https://github.com/milvus-io/milvus/discussions/48684#discussioncomment-16427291}} we instead adopt an HPC-oriented deployment that replaces MinIO with Lustre. In this configuration, all components directly access Lustre as a shared filesystem. To our knowledge, this setup is not publicly documented and represents a novel deployment approach for HPC environments.

We test each client implementation (e.g., Python, Go, Rust) and report only the results with the optimal client for brevity (Go for Weaviate and Milvus, and Rust for Qdrant). In this context we use the term ``worker'' to refer to an independent VDB node. However, Milvus relies on a set of microservices; for single-worker experiments, we use the closest equivalent: Milvus Standalone, a single-process deployment of its architecture. During distributed testing, we place up to four workers per compute node to balance parallelism with node-level resource limits. Additionally, note that in all cases, clients are placed on a separate compute node from the VDB instance(s) to measure network performance and prevent resource contention.

\subsection{VECHINI}
To support reproducible VDB evaluation in HPC environments, we design \texttt{VECHINI}: \textbf{V}ector Database \textbf{E}valuation and \textbf{C}haracterization for \textbf{H}PC \textbf{I}nsertion, Indexing, and \textbf{N}earest-neighbor Search. Existing VDB benchmarking frameworks~\cite{aumuller2018annbenchmarksbenchmarkingtoolapproximate, weaviate_ann_benchmark, qdrant_benchmarks, milvus_vdbbench_2025, simhadri2022bigann} assume Docker- or Kubernetes-based orchestration, which is unavailable on many HPC systems. VECHINI instead uses HPC-native tools, including Apptainer and MPI, enabling practitioners to deploy VDBs within the security constraints of their HPC platforms. Using a schema-driven configuration layer, VECHINI generates experiments for common VDB tasks, including insertion, indexing, and querying. Experiments are defined using configuration files that the benchmark expands into experiment directories for submission to the HPC job scheduler. By modifying the configuration file, users can control key HPC features such as cores-per-worker, number of compute nodes, clients/workers-per-compute-node, and storage backend without managing system-specific implementation details. VECHINI currently supports Qdrant, Milvus, and Weaviate on the \polaris{} and \aurora{} systems; however, the framework is designed to be extensible, allowing users to add new VDBs and HPC platforms while reusing the same schema-driven configuration layer.

\subsection{Embedding Datasets}
Thoroughly evaluating VDBs requires accounting for varied embedding characteristics and dataset scale, as these jointly determine system behavior and performance~\cite{chen2026alayalaserefficientindexlayout, roofline2009}. Accordingly, we select two distinct small-scale embedding datasets, GIST~\cite{Jegou2011Product} and dbpedia-openai-1M~\cite{dbpedia-entities-openai-1M}, and two large-scale embedding datasets, Yandex-text-to-image (Yandex-T2I) and Pes2o-VE. GIST consists of $960$-dimensional embeddings derived from multiple image collections, while dbpedia-openai-1M~\cite{dbpedia-entities-openai-1M} is composed of $1560$-dimensional embeddings generated from text excerpts by the text-embedding-ada-0002~\cite{openai_embeddings_2022} model.



 \begin{table*}[t]
\centering
\vspace{5pt}
\caption{Cloud vs HPC vector database performance with a recall minimum of 0.95. All reported cloud results were obtained by a prior publicly available benchmarking effort by Qdrant\cite{qdrant_benchmarks} (denoted using ``\textit{cloud}" in the platform column). HPC results are the median value obtained from 3 runs. Note that Weaviate did not use deferred indexing in the original experiments and instead built the index during data ingestion, resulting in an index time of 0.}

\label{tab:vdb_cloud_hpc_performance}
\begin{tabular}{|l|l|l|c|c|c|r|r|r|r|r|}
\hline
\textbf{VDB} &
\textbf{Platform} &
\textbf{Dataset}  &
\textbf{M} &
\textbf{EF-C} &
\textbf{EF-S} &
\textbf{Recall} &
\textbf{Upload Time} &
\textbf{Index Time} &
\textbf{QPS} &
\textbf{Latency} \\
\hline








\multirow{4}{*}{Milvus}
 & \textit{Cloud} & GIST & 32 & 128 & 128 & 0.964 & 84.968 (s) & 487.061 (s) & 281.53 & 322.63 (ms) \\
 & \aurora{} & GIST & 32 & 128 & 128 & 0.961 & 197.29 (s) & 84.50 (s) & 1470.29 & 23.85 (ms) \\
 
 & \textit{Cloud} & dbpedia-openai & 32 & 128 & 128 & 0.998 & 176.66 (s) & 886.152 (s) & 154.06 & 582.51 (ms) \\
 & \aurora{} & dbpedia-openai & 32 & 128 & 128 & 0.996 & 129.03 (s) & 205.41 (s) & 799.50 & 46.87 (ms) \\
\hline

\multirow{4}{*}{Qdrant}
 & \textit{Cloud} & GIST & 32 & 512 & 256 & 0.960 & 146.29 (s) & 2014.81 (s) & 433.95 & 207.28 (ms) \\
 & \aurora{} & GIST & 32 & 512 & 256 & 0.980 & 45.16 (s) & 215.22 (s) & 1373.26 & 58.31 (ms) \\
 & \textit{Cloud} & dbpedia-openai & 16 & 128 & 64 & 0.967 & 238.42 (s) & 671.83 (s) & 1260.53 & 3.26 (ms) \\
 & \aurora{} & dbpedia-openai & 16 & 128 & 64 & 0.979 & 62.41 (s) & 85.12 (s) & 1738.55 & 3.24 (ms) \\
\hline

\multirow{4}{*}{Weaviate}
 & \textit{Cloud} & GIST & 64 & 512 & 512 & 0.970 & 836.96 (s) & N/A & 496.17 & 184.37 (ms) \\
 & \aurora{} & GIST & 64 & 512 & 512 & 0.988 & 281.84 (s) & N/A & 2481.86 & 30.72 (ms) \\
 & \textit{Cloud} & dbpedia-openai & 64 & 512 & 64 & 0.975 & 836.96 (s) & N/A & 1142.13 & 4.99 (ms) \\
 & \aurora{} & dbpedia-openai & 64 & 512 & 64 & 0.986 & 395.53 (s) & N/A & 1518.28 & 4.87 (ms) \\
\hline


\end{tabular}
\end{table*}


To expand our evaluation to larger datasets, we include Yandex-T2I~\cite{Yandex2016_Deep1B}, a 745~GB multimodal dataset for cross-modal retrieval. Additionally, we introduce \texttt{Pes2o-vector-embeddings} (Pes2o-VE). Pes2o-VE is motivated by a biology application that augments BV-BRC ~\cite{BVBRC}, a comprehensive bioinformatics resource, with context retrieved from an academic corpus. Using 22,723 genome-related terms, the VDB is queried for relevant passages, which---along with structured records---are provided to a scientific reasoning model to generate structured “bio-narratives” for a downstream RAG pipeline. Pes2o-VE is generated from the Pes2o text corpus~\cite{peS2o} (8M+ academic papers) using a combination of recursive splitting and semantic chunking with the Qwen3-Embedding-4B model~\cite{qwen3embedding}. The resulting dataset contains 88M embeddings (843.56~GB), comparable in scale to many of the largest available embedding benchmarks (e.g., Yandex-T2I, LAION-400M \cite{schuhmann2021laion400mopendatasetclipfiltered}).

\subsection{Selecting Representative VDB Workload Patterns}

We consider two common patterns observed in VDB workloads: insert-then-query (a bulk write phase followed by a sustained read phase) and mixed insert-query (interleaved read and write operations). The former pattern is common to knowledge retrieval tasks such as documentation search~\cite{Salsabilla2025_Implementation}, case-law review~\cite{MENTZINGEN2024100247,harvard_lil_cold_cases_dataset}, and traditional RAG pipelines~\cite{singh2025ai2scholarqaorganized,lo-etal-2020-s2orc,lewisRetrievalAugmentedGenerationKnowledgeIntensive,xia2025multimodalfoundationmodelcosmological,Ji2024Automating}, while the latter is characteristic of emerging agentic workloads~\cite{maharana2024evaluatinglongtermconversationalmemory,argonne_resilience_ai_assistant} and online-content recommendation systems~\cite{Covington2016_youtube}. Notably, in practice, even mixed insert-query workloads typically begin with an initial phase that inserts background knowledge (e.g., relevant scientific literature for an agent's task), highlighting the importance of the initial ingestion step.



\subsection{Experimental Metrics}
\label{sec:expMetrics}
Throughout our experiments, we use the following terminology to describe key aspects of VDB workloads: 
\begin{itemize}
\item \textbf{Upload (insert) time}: Time to transfer data from client(s) to the VDB, measured as time to ``searchability,'' namely, when newly inserted data becomes visible to query.

\item \textbf{Index time}: Time to construct the index over all data and reach a steady state with no further indexing. 

\item \textbf{Queries per second (QPS)}: Number of queries completed per second.

\item \textbf{Query latency}: End-to-end time to complete a query or query batch.

\item \textbf{P95/99 latency}: Latency below which a given threshold (95/99\%) of query or query batches complete, capturing tail behavior.

\item \textbf{Recall}: The fraction of true nearest neighbors retrieved relative to exhaustive search (see \Cref{eq:recall_at_k}).
\end{itemize}

\section{From Cloud to HPC: Performance Impact}
\label{sec:cloud_v_hpc}
To study key VDB performance differences between cloud and HPC settings, we execute an open-source cloud benchmark suite ~\cite{qdrant_benchmarks} on \aurora{}. 
We employ identical parameters and versions as stated in the published benchmark results. \footnote{\url{https://github.com/qdrant/vector-db-benchmark/tree/c5b4d45659feafaaa968b6f07fdc12a7eb20e171}} The original experiments were conducted using two nodes: a client node with 8 vCPUs, 16 GiB of memory, and 64 GiB of storage running 100 Python clients; and a server node for the VDB with 8 vCPUs, 32 GiB of memory, and 64 GiB of storage. During execution, each Python client sends sequential non-batched queries, operating in parallel. We mimic this deployment model, replacing the cloud instances with \aurora{} compute nodes described in \Cref{tab:systems}. To select configurations for testing, we filter by a minimum recall threshold, using a cutoff of 0.95 recall to reflect the original constraints of the benchmark. For settings that meet the minimum recall requirement, we sort configurations by QPS and select the best-performing configuration.



\subsection{Imbalanced Impact of HPC Resources}
As shown in \Cref{tab:vdb_cloud_hpc_performance}, the benchmarks exhibit markedly different performance characteristics in cloud environments than in HPC, with all VDBs exhibiting substantial increases in QPS (1.33$\times$--5.22$\times$) and corresponding reductions in latency (1.01$\times$--13.53$\times$). Interestingly, we observe diminishing returns for latency, where beyond a certain point, additional computational resources no longer meaningfully reduce response time. The dbpedia-openai results for Qdrant and Weaviate illustrate this effect clearly: although QPS increases on HPC hardware, individual query latency remains largely unchanged. Additionally, the magnitude of improvement varies across VDB systems. Milvus, in particular, demonstrates the largest relative gains on HPC hardware across both datasets, reducing latency by an average of 12.98$\times$ and increasing throughput by an average of 5.21$\times$. 


Beyond query serving, HPC deployments also accelerate data ingestion and index construction. On average, upload time decreases by 2.32$\times$, while index construction time decreases by an average of 6.83$\times$. Notably, index build time exhibits the largest absolute reduction in execution time. This aligns with expectations as index construction is an inherently compute- and memory-intensive process, during which we typically observe sustained CPU utilization of $90\%+$. Consequently, HPC systems, with their higher core counts and increased memory bandwidth, are particularly well suited to accelerating index construction. In the majority of cases, upload time also decreases on HPC architectures; however, the gains are more modest (see \Cref{sec:IQ} for a root cause analysis). One notable outlier is the increase in upload time for GIST when using Milvus; additional HPC runs revealed variability in Milvus's GIST upload time (75--289 seconds). 





\subsection{Embedding Geometry and HPC Performance}
The benefit of HPC resources differs between the two datasets, with larger latency and QPS improvements observed for GIST. Prior work~\cite{aumuller2019rolelocalintrinsicdimensionality, Houle2018On,Elliott_2024} has shown that a metric known as intrinsic dimensionality (ID)~\cite{Levina2004Max} can serve as a proxy for search difficulty. ID measures the underlying complexity of the embedding space by estimating the minimum number of dimensions needed to describe its structure, where a higher ID indicates a more challenging search space. Analysis reveals that GIST has a higher ID (46.85) compared with dbpedia-openai (31.32). As a result, GIST requires a more intensive search to achieve 0.95 recall with Weaviate and Qdrant, thus benefiting more from the increased resources present in HPC. 

Our results reveal that VDB performance differs substantially between cloud and HPC systems. We observe uneven scaling across workflow stages and an interplay between acceleration and embedding geometry. These results expose fundamentally different performance properties, demonstrating that prior cloud-based evaluations do not directly generalize to HPC. 

\begin{lessonlist}
\begin{itemize}[leftmargin=*, label=, itemsep=2pt]
    \item \textbf{Lesson 1:} Indexing time and QPS significantly improve on HPC compared with the cloud, while already low-latency values remain unchanged.
    \item \textbf{Lesson 2:} Complex embedding spaces require greater search effort, increasing sensitivity to available compute and amplifying the impact of HPC resources.
\end{itemize}
\end{lessonlist}

\begin{figure}[t]
 \centering
  \includegraphics[width=\columnwidth]{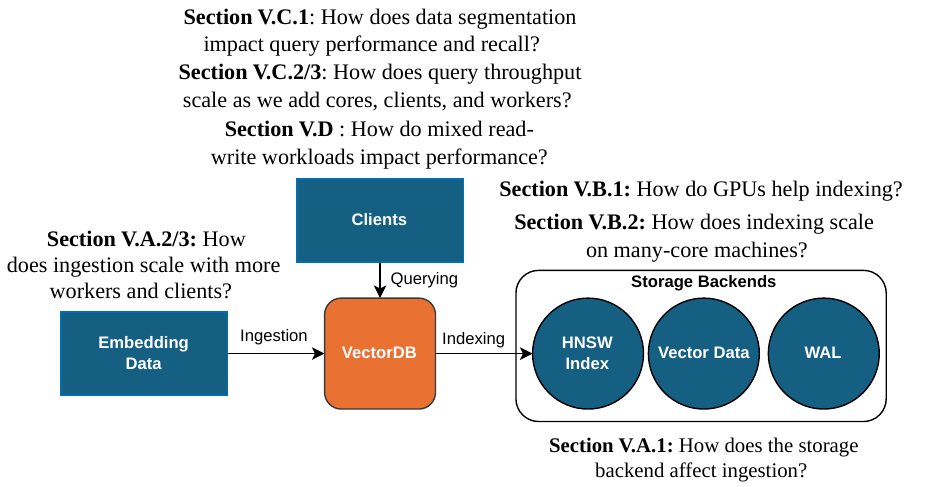}
\centering
\caption{Life cycle of a vector database. }
\label{fig:VDBLifecycle}
\vspace{-0.15in}
\end{figure}

\section{Evaluating the VDB Life Cycle on HPC Systems}
\label{sec:IQ}
 In the absence of VDB-focused studies on HPC systems, the community lacks a baseline for expected state-of-the-art performance and an understanding of how existing designs’ strengths and weaknesses manifest. To address this gap, we evaluate two representative workload patterns that span the spectrum of VDB usage in both industry and science: \textit{insert-then-query}~\cite{Salsabilla2025_Implementation,xia2025multimodalfoundationmodelcosmological,Ji2024Automating} and \textit{mixed insert-query}~\cite{maharana2024evaluatinglongtermconversationalmemory,argonne_resilience_ai_assistant,hellert2025ospreyproductionreadyagenticai}. \Cref{fig:VDBLifecycle} shows the typical life cycle of a VDB, which consists of data ingestion (upload), indexing, and querying, and key questions that we seek to answer in each section. Using the two workload patterns, we characterize how existing VDBs leverage HPC resources---such as multitier storage systems and GPU acceleration---and how they scale in many-core and distributed environments throughout their life cycle.


\subsection{Insertion}
In this section we evaluate the first stage of the VDB life cycle: ingestion/upload. \Cref{sec:storageMedium} examines how the underlying storage medium impacts performance, testing whether  VDBs can effectively leverage HPC’s multitier storage hierarchy. \Cref{sec:single_worker_multi_client_insert} evaluates the insertion limits of a single VDB worker to identify system-level bottlenecks that constrain throughput.  \Cref{sec:multiWorkerInsertion} explores distributed insertion scalability to determine whether existing VDBs can meet the demands of large-scale scientific datasets.

In a bulk upload scenario, queries occur after ingestion and indexing. To improve upload performance and align with developer recommendations,\footnote{\url{https://qdrant.tech/documentation/tutorials-develop/bulk-upload/}} we defer HNSW construction until after upload completion. During upload, all systems use a flat index (i.e., exhaustive search if queried). To measure ``searchability" (see \Cref{sec:expMetrics}), we utilize system-specific mechanims. For Milvus, we issue a sentinel query that succeeds once all entities are queryable; for Weaviate, we measure when the asynchronous indexing queue becomes empty; and for Qdrant, we use the count API with \texttt{exact=true} to confirm that all vectors are visible. For parity, we use a single shard per worker for all VDBs. To focus on large-scale, high-dimensional data characteristics of modern embeddings\cite{chen2024exploringmeaningfulnessnearestneighbor,muennighoff2023mtebmassivetextembedding,bhalla2025higherembeddingdimensioncreates}, we limit upload experiments to Pes2o-VE as a representative HPC embedding workload.



\begin{figure}[t]
 \centering
  \includegraphics[width=\columnwidth]{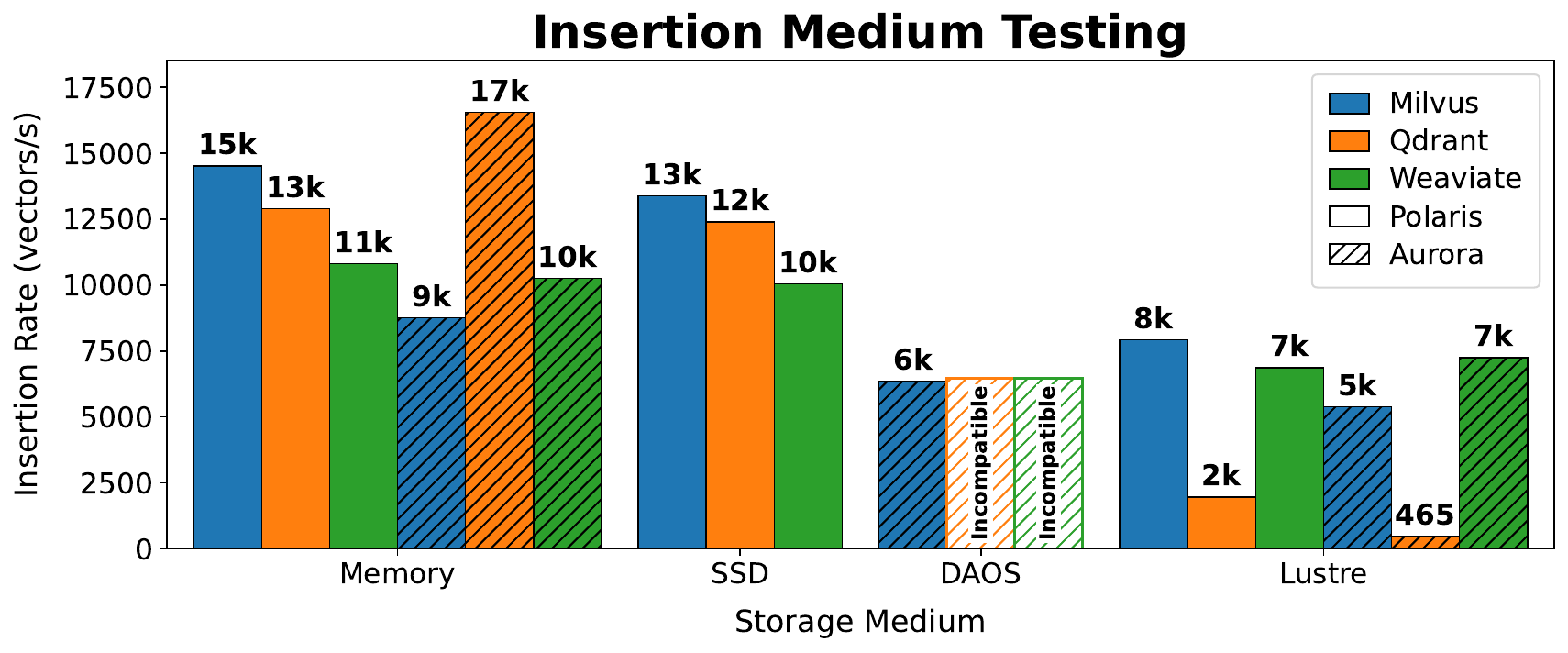}
 \caption{Pes2o-VE: Testing the impact of  different storage mediums  on insertion rate on \polaris{} and \aurora{}.}
 \label{fig:insertionMediumTesting}
\end{figure}
\subsubsection{Impact of Storage Medium on Performance}
\label{sec:storageMedium}
To understand the interaction between storage medium and insertion performance, we evaluate a variety of HPC storage backends using a single-client, single-worker setup. We sweep upload batch sizes from 32 to 32,768 vectors, selecting the best-performing option. We test each system’s supported storage backends: memory, SSD, and Lustre on \polaris{} and memory, DAOS, and Lustre on \aurora{}. To ensure computability with DAOS, we use the DFUSE POSIX interface \cite{Latham2025Initial}. We exclude Qdrant and Weaviate from DAOS experiments because several internal components of both VDBs utilize mmap system calls that are not supported by DAOS.

As shown in \Cref{fig:insertionMediumTesting}, the underlying storage medium has a substantial impact on achievable insertion throughput. Across all three VDBs, in-memory storage delivers the highest performance, followed by local SSD on \polaris{} and DAOS on \aurora{}, with Lustre consistently exhibiting the lowest throughput. Although Lustre is not designed for the small, frequent writes typical of ingestion, it remains necessary in situations that require long-term durability and significant storage capacity. Notably, DAOS achieves insertion rates approaching those of main memory with Milvus (8758 vectors/s vs 6432 vectors/s), highlighting its promise as an HPC-optimized persistence layer. 


As shown in \Cref{fig:insertionMediumTesting}, Qdrant insertion throughput drastically falls on Lustre compared with the other VDBs. Qdrant's performance is caused by the interaction between storage latency and its update pipeline. While all VDBs write to a write-ahead log (WAL) before asynchronously updating segments, Qdrant requires each insert to reserve a slot in a bounded update queue before completing the WAL write. This queue is drained by a thread applying segment mutations. Initially, reservation latency is negligible (0.000 ms); but because of slow segment mutation (1011 ms), the thread cannot drain the queue quickly enough. This results in significant queuing delay, with reservation latency rising to 1445 ms. In contrast, Lustre's latency impacts only a small portion of the critical write path in Milvus and Weaviate---primarily WAL persistence---resulting in lower performance degradation.

\begin{figure}[t]
 \centering
  \includegraphics[width=\columnwidth]{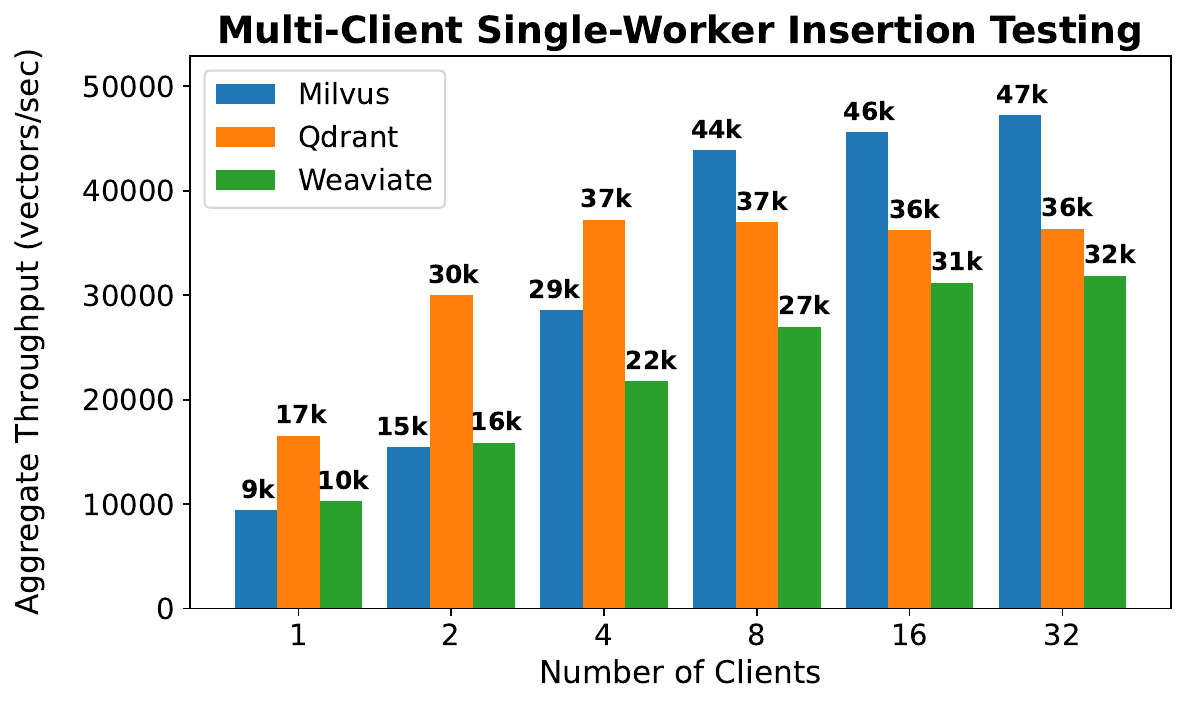}
\caption{Pes2o-VE: testing single-worker multiclient insertion on \aurora{}. }
\label{fig:insertTesting}
\vspace{-10pt}
\end{figure}

\begin{figure}[b]
 \centering
  \includegraphics[width=\columnwidth]{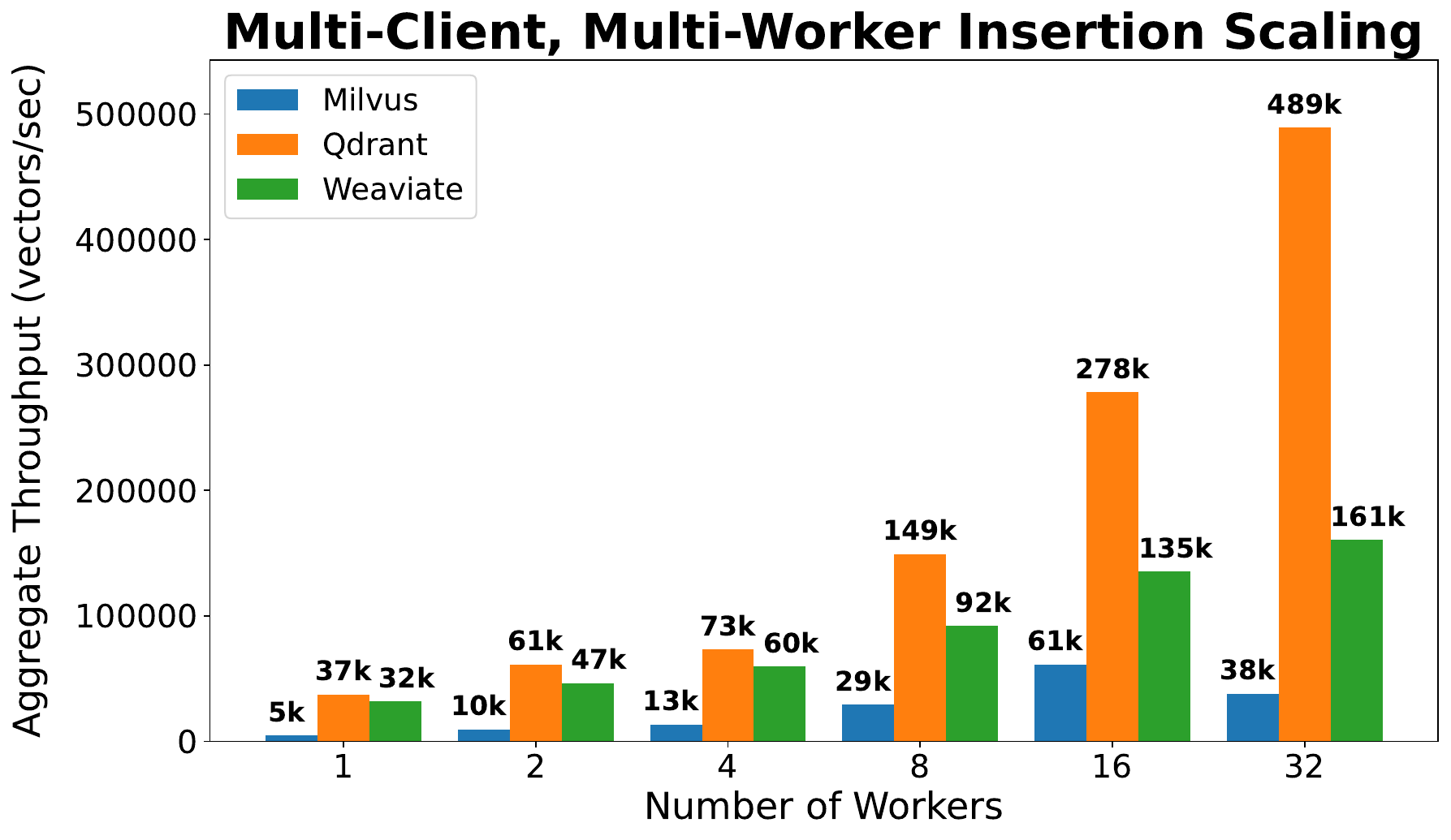}
\centering
\caption{Pes2o-VE: testing insertion scaling on \aurora{}.}
\label{fig:insertionScaling}
\vspace{-10pt}
\end{figure}


\subsubsection{Single-Worker Insertion}
\label{sec:single_worker_multi_client_insert}
To assess per-worker ingest scalability, we increase concurrent upload clients until performance saturates on \aurora{}. As shown in \Cref{fig:insertTesting}, Qdrant achieves the highest per-client throughput (37,218 vectors/s with four clients) but does not scale, with minimal gains beyond four clients due to severe WAL contention (104.034 ms waiting vs.\ 3.746 ms writing). Weaviate trails just behind Qdrant, peaking at 31,859 vectors/s with 32 clients. Milvus reaches a higher peak throughput (47,236 vectors/s at 32 clients) before declining, as insertion becomes dominated by the WAL append pipeline (67.3 ms of 75.0 ms), which saturates at higher concurrency. To demonstrate the impact of increased WAL parallelism, we modify Qdrant's and Milvus's configurations to increase the number of shards, focusing on Qdrant and Milvus because they achieved the highest insertion throughput. With 32 clients and WALs, Qdrant and Milvus achieve throughputs of 223,843 and 138,860 vectors/second, respectively. These results highlight the importance of WAL parallelism as a mechanism to increase resource utilization on HPC systems.

\subsubsection{Multiworker Insertion}
\label{sec:multiWorkerInsertion}
To support large-scale scientific datasets, VDBs must support efficient parallel ingestion. We evaluate multinode insert scaling by increasing distributed workers from 1 to 32 (powers of two). We use the optimal batch size and clients per worker from \Cref{sec:single_worker_multi_client_insert}, evenly partitioning data across workers to maximize performance. For Milvus, we utilize the HPC configuration described in \Cref{sec:methodology} and scale the components responsible for initial data ingestion: proxies and streaming nodes. 

As shown in \Cref{fig:insertionScaling}, Weaviate and Qdrant's insertion rate increases as the number of workers increases, achieving   160,551 vectors/s and 489,213 vectors/s, respectively. However, both systems scale sublinearly. Milvus’s insertion rate scales up to a maximum of 61,344 vectors/second with 16 proxies and streaming nodes. Beyond that, throughput decreases. As an alternative to streaming-based ingestion, we also evaluated Milvus’s bulk insertion utility, which bypasses the WAL and writes directly to object storage. Utilizing memory-backed MinIO-storage to maximize performance, we achieved an upload rate of 83,042 vectors/second. 

\subsubsection{Discussion}
Notably, even the best-performing VDB's insertion throughput is well below the hardware capabilities of the HPC interconnect, which supports multiple 25 GB/s links (a theoretical maximum of roughly 2.6 million Pes2o-VE vectors per second per link) Although server-side components of the insertion pipeline (e.g., WAL writes, locking) are expected to limit achievable bandwidth, the observed gap suggests that enforcing per-insert consistency may be mismatched to the initial bulk ingestion phase of many VDB workloads. In this setting, such guarantees can unnecessarily constrain throughput. This observation motivates the need for ingestion modes that relax consistency guarantees during bulk loading to better utilize available system resources. Additionally, we observe that write amplification is a nontrivial cost. For example, a 95~GB subset of Pes2o-VE expands to 266~GB on Lustre using Milvus---an overhead that is tolerable at small scale but becomes prohibitive for large-scale workloads.


\begin{lessonlist}
\begin{itemize}[leftmargin=*, label=, itemsep=2pt]
    \item \textbf{Lesson 3:}  Persistent storage latency dominates insertion performance: WAL latency and Lustre backpressure prevent VDBs from fully utilizing available bandwidth.

    \item \textbf{Lesson 4:} HPC VDB deployments must treat write amplification as a key optimization metric, because additional storage overhead will quickly become intractable at scale.
\end{itemize}
\end{lessonlist}

\begin{table}[t]
\centering
\caption{Mean indexing time (seconds) across vector databases, methods, and HPC systems. Experiments were repeated three times; we observed minimal variance, and therefore omit standard deviation from the table.}
\label{tab:cpu_gpu_indexing}
\begin{tabular}{|l|ll|cc|cc|}
\hline
\textbf{VDB} & \textbf{Dataset} & \textbf{Size} 
& \multicolumn{2}{c|}{\textbf{SC1} } 
& \multicolumn{2}{c|}{\textbf{SC2 } }\\
\hline
 & & 
 & CPU & GPU 
 & CPU & GPU \\
\hline

\multirow{6}{*}{Milvus}
& \multirow{3}{*}{Pes2o-VE}
& 1M   & 485 & 143 & 119 & N/A \\
& & 5M  & 2398 & 718 & 368 &  N/A \\
& & 10M &  4818 & 1541 & 691 &  N/A \\
\cline{2-7}
& \multirow{3}{*}{Yandex-T2I}
& 1M   &  91 & 61 & 37 & N/A \\
& & 5M  &  223 & 96 & 78 &  N/A \\
& & 10M &  464 & 163 & 110 &  N/A \\
\hline

\multirow{6}{*}{Qdrant}
& \multirow{3}{*}{Pes2o-VE}
& 1M   & 160 & 15 & 73 &  N/A \\
& & 5M  & 1077 & 61 & 420 &  N/A \\
& & 10M & 1847 & 109 & 845 &  N/A \\
\cline{2-7}
& \multirow{3}{*}{Yandex-T2I}
& 1M   & 20 & 5 & 8 &  N/A \\
& & 5M  & 109 & 22 & 58 &  N/A \\
& & 10M & 219 & 45 & 131 &  N/A \\
\hline

\multirow{6}{*}{Weaviate}
& \multirow{3}{*}{Pes2o-VE}
& 1M   & 1176 & N/A &  1726 &  N/A \\
& & 5M  & 6294 & N/A & 9779 &  N/A \\
& & 10M & 13,402 & N/A & 19,457 &  N/A \\
\cline{2-7}
& \multirow{3}{*}{Yandex-T2I}
& 1M   & 498 & N/A & 487 &  N/A \\
& & 5M  & 3001 & N/A & 3120 &  N/A \\
& & 10M & 6415 & N/A &  5951 &  N/A \\
\hline


\hline

\end{tabular}
\end{table}









\subsection{Indexing}
After data ingestion, each VDB advances to the next stage of the VDB lifecycle---HNSW indexing---where we examine how HPC resources influence performance. \Cref{sec:cpuGPUIndexing} studies the impact of GPU acceleration on indexing, since GPUs are a fundamental component of HPC systems. \Cref{sec:index_core_testing} evaluates the effect of many-core CPUs on indexing to expose scaling limitations. For clarity, note that we use the term \textit{cores} to refer to the physical cores on a compute node's CPU, while we use the term \textit{virtual cores} to refer to the node's hardware threads (e.g., \aurora{} has 104 cores and 208 virtual cores). 

We select  $M=16$ and EF-C=100 as our HNSW parameters because they are commonly used baseline configurations in both the canonical literature~\cite{malkov2018efficientrobustapproximatenearest} and popular ANN implementations~\cite{hnswlib,opensearch_knn_docs}, providing a balanced trade-off between memory usage, construction cost, and recall. Note that for GPU-accelerated indexing, Milvus lacks a direct equivalent to HNSW. As a proxy, we employ its graph-based GPU-CAGRA index and configure its hyperparameters ($M=16$, intermediate node edge limit during construction $=64$) to approximate a similar construction process. Additionally, Weaviate does not provide GPU acceleration and is therefore excluded from GPU-based testing.  Note that we omit reporting multinode index testing because it is an embarrassingly parallel workflow, and we observed near-linear scaling using Qdrant with the full Pes2o-VE dataset. 


\subsubsection{Cross-System CPU-GPU Indexing Analysis}
\label{sec:cpuGPUIndexing}
We begin by evaluating indexing efficiency and the impact of GPU acceleration across varying data scales (1 million, 5 million, and 10 million) on \polaris{} and \aurora{} with Peso2-VE and Yandex-T2I.  \Cref{tab:cpu_gpu_indexing} presents the results, with each experiment repeated three times and summarized using the mean. Our results show that indexing time is strongly influenced by both dataset size and embedding dimensionality. The Pes2o-VE dataset (2560 dimensions) consistently incurs significantly higher indexing costs than does the lower-dimensional Yandex-T2I dataset (200 dimensions). This trend is especially relevant because modern scientific and multimodal workloads increasingly rely on higher-dimensional embeddings. Interestingly, GPU acceleration exhibits a similar trend. Although it improves indexing performance across both datasets, at 10M points  the gains are substantially larger for Pes2o-VE (10.04$\times$) than for Yandex-T2I (3.86$\times$), reflecting the increased computational intensity of higher-dimensional vectors and their suitability for GPU-based parallelism.

Focusing on CPU performance, \aurora{}’s 104-core nodes reduce indexing time compared with  \polaris{}'s 32-core nodes, by an average of 4.08$\times$ and 2.59$\times$ for Pes2o-VE-10M and Yandex-T2I-10M with Qdrant and Milvus, respectively. On \polaris, Qdrant is consistently the fastest, whereas Milvus outperforms Qdrant on \aurora{}, and Weaviate trails both VDBs by a significant margin on both platforms. The divergence in indexing time across VDBs highlights fundamentally different scaling behaviors, which we explore further in subsequent analysis.



\begin{lessonlist}
\begin{itemize}[leftmargin=*, label=, itemsep=2pt]
    \item \textbf{Lesson 5:} GPU-accelerated indexing delivers the greatest gains on high-dimensional datasets, where increased computational intensity and memory bandwidth demands better align with GPU architectures.
\end{itemize}
\end{lessonlist}

\subsubsection{Impact of Core Count on Indexing}
\label{sec:index_core_testing}
\begin{figure}[t]
 \centering
  \includegraphics[width=\columnwidth]{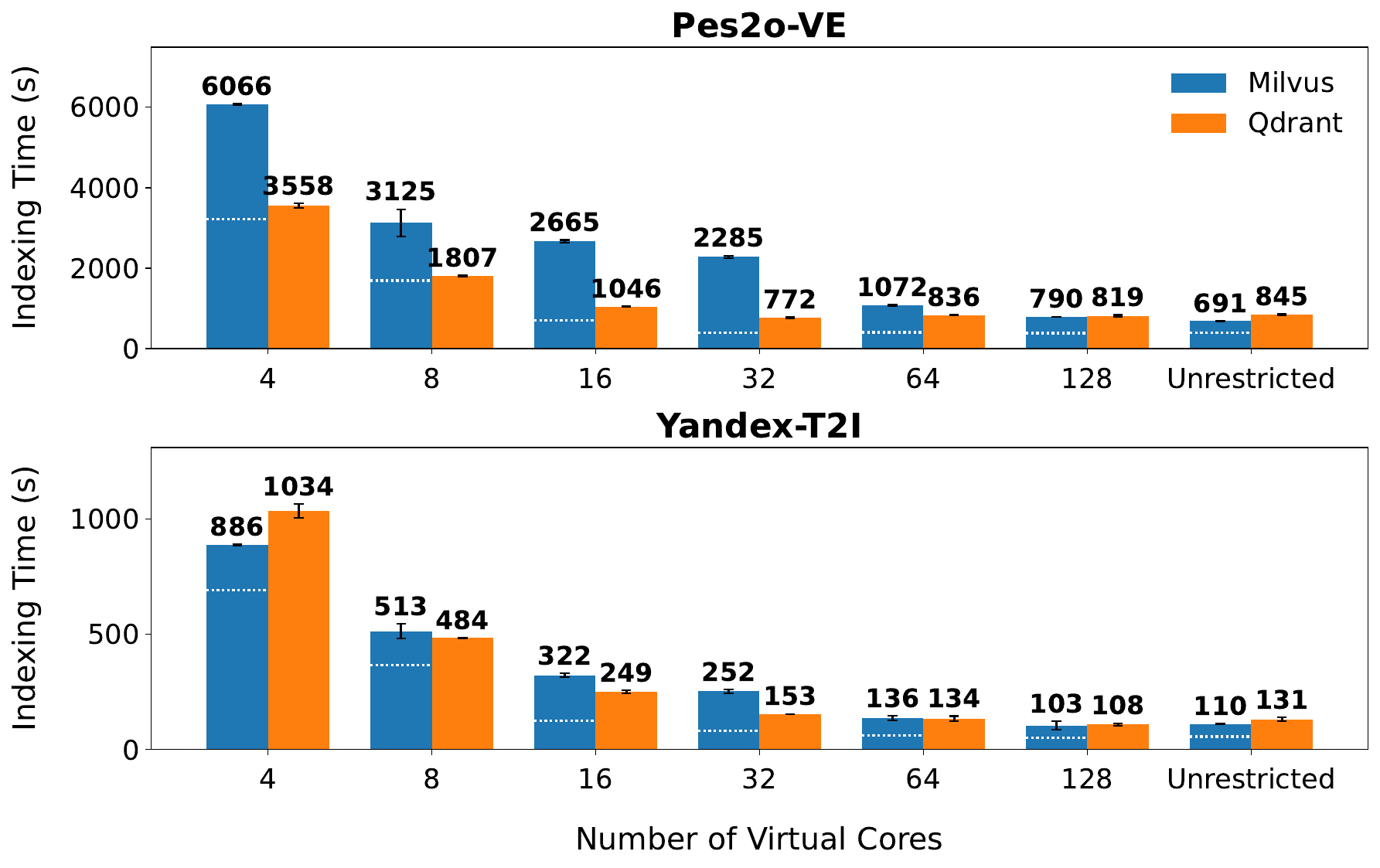}
\centering
\caption{Pes2o-VE-10M and Yandex-T2I=10M index core testing on \aurora{}. The white line  in the bars indicates when Milvus finishes constructing an initial index. Weaviate is omitted because its index-building process is sequential. }
\label{fig:index_core_testing}
\vspace{-8pt}
\end{figure}

This section examines how indexing performance scales with virtual core count on \aurora{}. Using MPI-based binding, we restrict available virtual cores in powers of two, with an additional unrestricted configuration for execution without any binding. For Milvus and Weavite, we limit internal parallelism to match the number of virtual cores using \texttt{GOMAXPROCS}, while Qdrant automatically adjusts its thread-count to match the selected binding. To control compute-hour costs, we use a 10M subset of each dataset. Experiments start at 4 virtual cores, because smaller configurations resulted in timeouts that triggered runtime failures. Notably, Weaviate’s transition from a flat to HNSW index is implemented as a batched sequential loop,\footnote{\url{https://github.com/weaviate/weaviate/blob/v1.36.0/adapters/repos/db/vector/dynamic/index.go#L631}} where each point in the batch is also inserted sequentially.\footnote{\url{https://github.com/weaviate/weaviate/blob/v1.36.0/adapters/repos/db/vector/hnsw/insert.go#L224}} As a result, this phase does not benefit from additional cores, and we therefore omit Weaviate from \Cref{fig:index_core_testing} and discuss its indexing behavior separately in the text.

As shown in \Cref{fig:index_core_testing}, increasing the number of virtual cores reduces indexing time for both Milvus and Qdrant, but with clear diminishing returns. For both datasets, the majority of gains are realized by 32--64 virtual cores, after which improvements taper off or reverse. Qdrant scales effectively at lower virtual core counts, achieving significant reductions in indexing time initially but limited gains past 32--64 virtual cores. This behavior is likely connected to its internal optimizer, which dynamically adjusts segment counts and indexing threads based on the detected virtual core count, creating varying thread-per-segment trade-offs. These dynamics are dataset dependent and warrant further study to fully understand their impact.

Milvus benefits from additional resources on Pes2o-VE, while on Yandex-T2I its performance peaks around 128 virtual cores. Notably, Milvus often completes its initial index faster than both Qdrant and Weaviate. However, because of the stabilization phase, during which background optimizations continue after the initial index is built, it falls beyond Qdrant with less than 128 and 64 virtual virtual cores with Pes2o-VE and Yandex-T2I, respectively. This creates a trade-off: while the initial index can be used to reduce wait time, we observe up to a 9.6$\times$ decrease in query throughput with the Pes2o-VE dataset when using the unoptimized index.

Weaviate’s comparatively longer indexing time is a consequence of its internal indexing architecture. In dynamic index mode, Weaviate initially inserts vectors into a flat index and upgrades the collection to HNSW in a sequential loop once a threshold is reached. Moreover, because Weaviate maintains a single index per shard, it cannot build multiple indexes in parallel within a shard. This limits its ability to exploit the many-core architecture of modern HPC nodes and causes its indexing performance to lag behind the other VDBs. We selected Weaviate's dynamic index because it mirrors the design of both Qdrant and Weaviate; however, we also evaluate Weaviate’s asynchronous indexing mechanism, which builds the HNSW index concurrently with data insertion using multiple threads. On the Yandex-T2I dataset, Weaviate's asynchronous indexing substantially reduced the total time to complete insertion and indexing with 10 million vectors from 6510 seconds to 709 seconds with comparable recall (0.7394 vs 0.738) and query throughput (8297 vs 7662 QPS).








\begin{figure*}[t]
 \centering
  \includegraphics[scale=0.32]{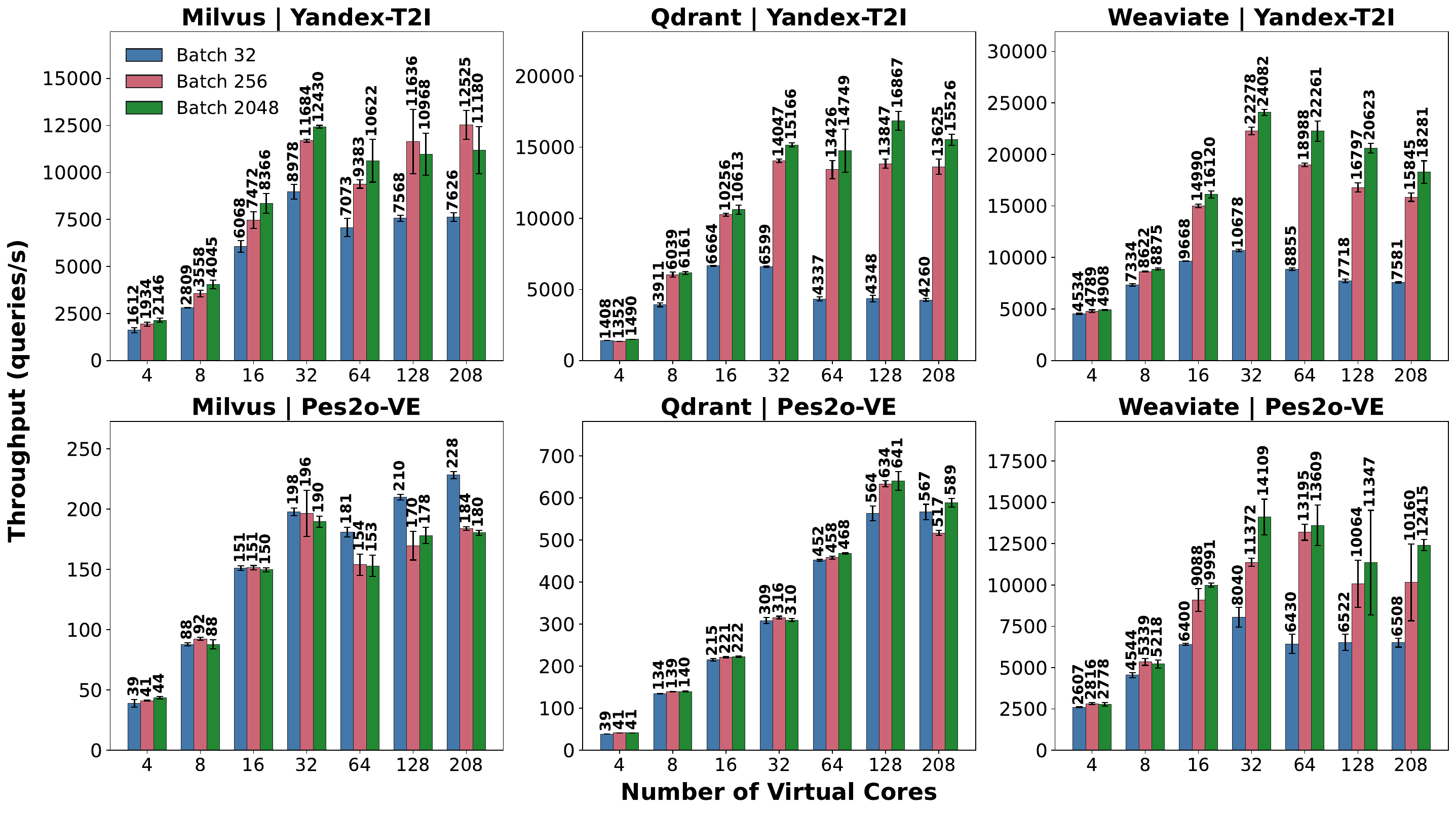}
\centering
\caption{Query performance with varied cores for Yandex-T2I-10M and Pes2o-VE-10M on \aurora{}.}
\label{fig:queryCoreTesting}
\end{figure*}

\subsection{Query}
\label{sec:query}
We now transition to the final stage of the VDB life cycle: querying. \Cref{sec:recallTesting} examines hidden query recall-cost trade-offs introduced by the common practice of data segmentation, while \cref{sec:coreScaling} tests single-node scalability, varying the number of virtual cores available. \Cref{sec:multiWorkerQuery} assesses multinode scalability, testing with up to 256 workers on 64 compute nodes. \Cref{sec:mixedWorkloads} examines how performance changes when insertion and querying occur simultaneously. We evaluate an in-memory search scenario in which all data is loaded prior to query execution, and we place the VDBs on a memory-backed file system. Because Weaviate's Go client does not provide a method to issue batch queries, we issue batches of independent single-query requests using concurrent Goroutines. Unless otherwise noted, queries are sent sequentially in batches, and we set $efSearch = 64$, a commonly used mid-range HNSW value~\cite{malkov2018efficientrobustapproximatenearest,aumuller2018annbenchmarksbenchmarkingtoolapproximate} that balances latency and recall.

\begin{table}[t]
\centering
\caption{Recall@10 comparison across VDBs for Yandex-T2I-10M and Pes2o-EV-10M datasets using full node-resources.}
\label{tab:recall-comparison}
\begin{tabular}{|l|c|c|}
\hline
\textbf{VDB} & \textbf{Yandex-T2I-10M} & \textbf{Pes2o-EV-10M} \\
\hline
Weaviate & 0.738 &  0.918 \\
\hline
Qdrant   & 0.837 & 0.970 \\
\hline
Milvus   & 0.876 & 0.982 \\
\hline
\end{tabular}
\vspace{-10pt}
\end{table}

\subsubsection{Single-Node Query Recall} 
\label{sec:recallTesting}
\Cref{tab:recall-comparison} provides a comparison of each VDB's mean recall across three experiments using all available cores, a batch size of 32, and 10M subsets of Pes2o-VE and Yandex-T2I. Notably, even with identical index parameters, the VDBs achieve different recall values: Milvus achieves the highest recall on Yandex-T2I, while Qdrant and Milvus achieve similar recall on Pes2o-VE. In both cases, Weaviate trails substantially behind the other two systems. 

The gap between Milvus/Qdrant and Weaviate is caused by the two different approaches to single-shard indexing. Weaviate maintains a single HNSW index per shard, meaning the graph is constructed over the entire shard. In contrast, Qdrant and Milvus partition each shard into multiple segments, where each segment maintains an independent index that captures local data patterns. During search, Qdrant and Milvus query these segment-level HNSW graphs in parallel and then merge the segment-level top-$k$ results into a final shard-wide top-$k$ through reranking. In effect, this increases the thoroughness of the search within each subspace and expands the candidate set considered during reranking, improving recall at the cost of increased computation per query.

\begin{figure*}[t]
 \centering
  \includegraphics[scale=0.35]{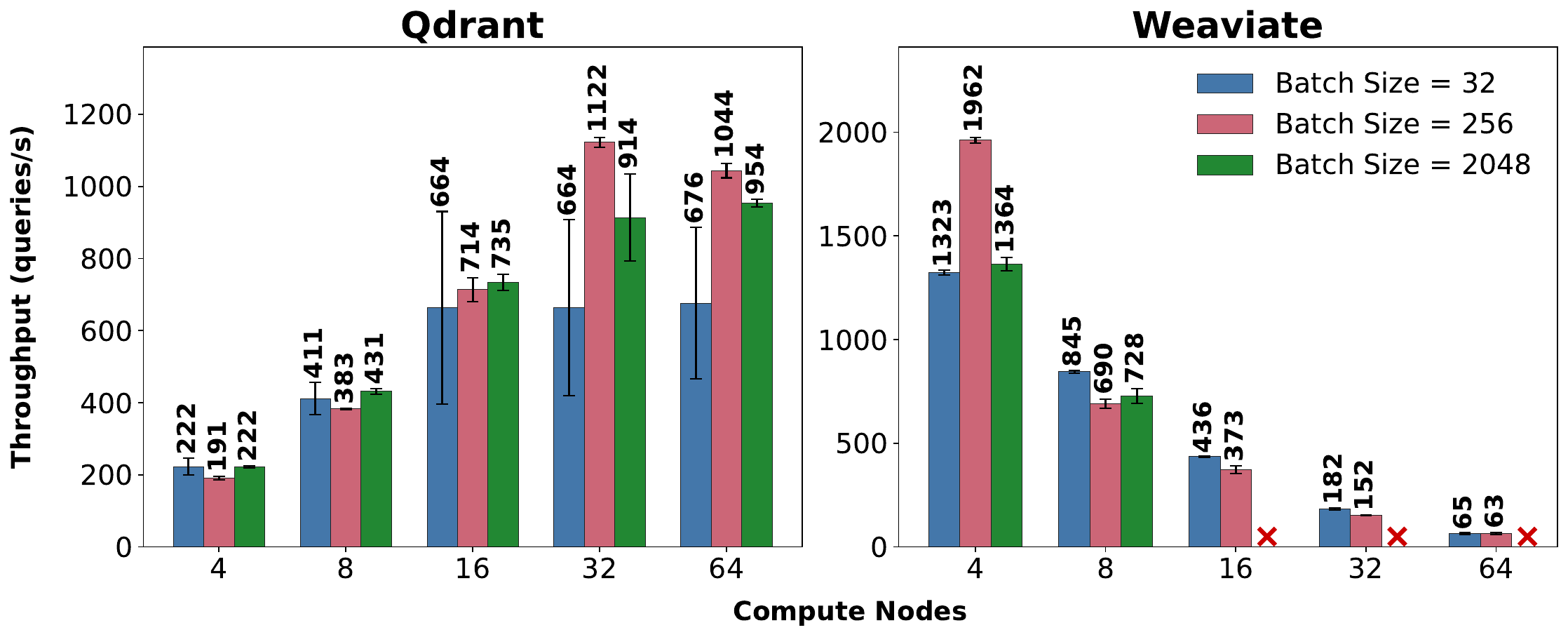}
\caption{Distributed query testing with the full Pes2o-VE dataset on \aurora{}. Note that because of runtime failures while using Lustre as their storage backend, we were unable to collect Milvus results (see \Cref{sec:multiWorkerQuery}). Each red ``X" in the Weaviate graph indicates configurations where query testing repeatedly failed because of Weaviate intracluster networking failures. }
\label{fig:multiNodeQuery}
\end{figure*}

Follow-up experiments with Qdrant using a 1M subset of Pes2o-VE confirmed our hypothesis: increasing the number of segments from 1 to 8 improved recall from 0.962 to 0.9854 but also increased total search time from 4.19 s to 22.65s. Additionally, with full-node resources, we find that Qdrant and Milvus use different numbers of segments with Yandex-T2I-10M (8 vs 5), while with Pes2o-VE-10M, Milvus and Qdrant use 103 and 25 segments, respectively. In both cases, the VDB using more segments achieved higher recall; however, the Pes2o-VE results suggest that beyond some point, additional segments provide diminishing returns. These results show that each VDB encodes a largely unstudied cost-recall trade-off that extends beyond the standard HNSW parameters, motivating further investigation into optimal segment sizes. 



\begin{lessonlist}
\begin{itemize}[leftmargin=*, label=, itemsep=2pt]
    \item \textbf{Lesson 6:} Modern VDBs make implicit cost-recall trade-offs by splitting large datasets into multiple segments, each with its own index. Utilizing more segments generally results in higher recall at the cost of increased computation per query. 
\end{itemize}
\end{lessonlist}



\subsubsection{Single-Node Query Core Scaling} 
\label{sec:coreScaling}
To explore single-node query scaling, we vary virtual cores from 4 to the full node resources by powers of two in the same manner described in \Cref{sec:index_core_testing}. We use 10M-vector subsets for tractability and evaluate batch sizes of 32, 256, and 2048, repeating all experiments three times to calculate mean and standard deviation. 
\paragraph{Milvus}
\Cref{fig:queryCoreTesting} shows that Milvus exhibits inconsistent scaling behavior: in 4/6 configurations, Milvus reaches its peak performance with only 32 of the 208 available virtual cores. In all configurations, performance shows diminishing returns beyond this point, with the per-core marginal gain dropping substantially. Perf counters indicate this is due to increasing memory pressure. As the number of cores increases, memory stall cycles grow from 65.3\% to 85.23\%, leading to a drop in IPC (0.54→0.14) and retiring cycles (9.1\%→ 3.6\%). 


\paragraph{Qdrant}
Qdrant’s scaling behavior appears more conventional at first glance, particularly with Pes2o-VE. However, Qdrant's scaling behavior is largely governed by internal optimization policies. Qdrant increases its internal search threads based on the detected core count and, inversely, changes settings that can cause the number of segments to decrease with more cores. This creates an effect where at lower core counts, there are significancy fewer segments and less threads-per-segment with Pes2o-VE, increasing search latency. For example, searching 50 segments with 32 virtual cores takes 98 ms, compared with 51 ms for 26 segments with 208 virtual cores (batch size = 32). In scenarios where the number of segments largely does not change across core counts---such as with the smaller Yandex-T2I dataset ---we observe that scaling tapers after 32 cores. We see similar limited scaling behavior past 32 cores with Pes2o-VE when we fix the number of segments to 26 (the default selected with full node resources). This demonstrates that while Qdrant’s segmentation model enables dynamic adaptation, it does not enable Qdrant to fully take advantage of many-core architectures.
\paragraph{Weaviate}
Across batch sizes and datasets, Weaviate achieves its highest QPS by 32 virtual cores before its performance decreases with more resources. As shown in our Milvus analysis, HNSW traversal is memory-bound, with concurrent traversals contending for shared in-memory data structures. Furthermore, Weaviate's architectural design limits parallelism to the batch level. Unlike Qdrant and Milvus, which partition each shard and utilize multiple HNSW graphs, Weaviate maintains a single HNSW graph per shard, further limiting parallelism.
 
\begin{lessonlist}
\begin{itemize}[leftmargin=*, label=, itemsep=2pt]
    \item \textbf{Lesson 7:} QPS improves up to 32 cores but often plateaus or degrades beyond that, highlighting that current VDBs do not scale effectively on many-core HPC systems.
\end{itemize}
\end{lessonlist}

\begin{table}[t]
\centering
\caption{Mean Recall@10 with the full Pes2o-VE dataset and a query batch size of 32 across compute-node counts for Qdrant and Weaviate.}
\label{tab:recall-vdb-nodes}
\begin{tabular}{|r|c|c|}
\hline
\textbf{Compute Nodes} & \textbf{Qdrant} & \textbf{Weaviate} \\
\hline
4  & 0.946 & 0.910 \\
\hline
8  & 0.954 & 0.920 \\
\hline
16 & 0.962 & 0.934 \\
\hline
32 & 0.943 & 0.935 \\
\hline
64 & 0.959 & 0.931 \\
\hline
\end{tabular}
\end{table}

\begin{table*}[t]
  \centering
  \caption{Concurrent update and query performance. QPS = query vectors/second. Batch
  query latency and P99 latency are reported in milliseconds. }
  \label{tab:mixedQueryInsert}
  \setlength{\tabcolsep}{3.2pt}
  \begin{tabular}{|c|c|c|c|c|c|}
  \hline
  \textbf{Data} & \textbf{System} & \textbf{Pattern} & \textbf{QPS} & \textbf{Mean Latency}
  & \textbf{P99 Latency} \\
  \hline
  \multirow{6}{*}{Pes2o}
& Milvus   & Baseline & $442.49 \pm 4.34$ & $72.23 \pm 0.71$ & $99.50 \pm 2.46$ \\
  \cline{2-6}
  & Milvus   & Mixed    & $395.54 \pm 8.69$ & $81.05 \pm 1.81$ & $145.16 \pm 17.51$ \\
  \cline{2-6}
  
  & Weaviate & Baseline & $7{,}853.42 \pm 151.47$ & $4.07 \pm 0.08$ & $5.57 \pm 0.11$ \\
  \cline{2-6}
  & Weaviate & Mixed    & $4{,}430.65 \pm 248.48$ & $7.41 \pm 0.42$ & $36.41 \pm 1.56$ \\
  \cline{2-6}
  & Qdrant   & Baseline & $751.42 \pm 25.13$    & $42.56 \pm 1.40$ & $59.15 \pm 1.50$ \\
  \cline{2-6}
  & Qdrant   & Mixed    & $355.13 \pm 13.33$    & $90.12 \pm 3.45$ & $201.91 \pm 20.18$ \\
  \hline
  \multirow{6}{*}{Yandex-T2I}
& Milvus   & Baseline & $11{,}812.39 \pm 55.47$ & $2.71 \pm 0.01$ & $3.63 \pm 0.20$ \\
  \cline{2-6}
  & Milvus   & Mixed    & $7{,}528.91 \pm 301.91$ & $4.27 \pm 0.17$ & $21.27 \pm 0.26$ \\
  \cline{2-6}

  & Weaviate & Baseline & $10{,}671.77 \pm 136.08$ & $3.00 \pm 0.04$ & $4.84 \pm 0.14$ \\
  \cline{2-6}
  & Weaviate & Mixed    & $7{,}132.85 \pm 228.76$  & $4.53 \pm 0.14$ & $14.95 \pm 0.47$ \\
  \cline{2-6}
  & Qdrant   & Baseline & $4{,}754.52 \pm 227.67$  & $6.74 \pm 0.33$ & $8.25 \pm 1.44$ \\
  \cline{2-6}
  & Qdrant   & Mixed    & $2{,}453.64 \pm 129.63$  & $13.05 \pm 0.71$ & $36.24 \pm 4.53$ \\
  \hline
  \end{tabular}
  \vspace{-10pt}
\end{table*}

\subsubsection{Multiworker Testing} 
\label{sec:multiWorkerQuery}
After evaluating single-node scalability, we examine multinode strong scaling using the full Pes2o-VE dataset, scaling up to 64 compute nodes and testing batch sizes of 32, 256, and 2048. We use a single client that issues sequential batches, keeping the workload fixed while increasing the number of workers. To reduce compute-hour cost while still utilizing real-world scientific embeddings, we focus on the larger, higher-dimensional dataset Pes2o-VE. Note that while using a batch size of 2048 with more than 8 compute nodes, Weaviate's query operations repeatedly failed because of connection timeouts while contacting remote shards, preventing the collection of these results. Additionally, as will be discussed shortly, we were unable to scale Milvus to the full Peso2-VE dataset. All experiments are repeated three times, and we report the mean performance.



\Cref{fig:multiNodeQuery} presents the results of scaling from 4 to 64 compute nodes, while \Cref{tab:recall-vdb-nodes} reports the achieved recall. Despite using identical HNSW parameters, the VDBs again exhibit a  cost-recall trade-off, with Qdrant achieving slightly lower throughput but higher recall than Weaviate. Additionally, we observe that recall generally improves as the number of compute nodes increases. This mirrors the effect observed with segmentation: distributing the data across more compute nodes produces more independent HNSW graphs, improving candidate selection and increasing the number of candidates merged during the final reduction.  Unlike increasing the number of local segments, however, the additional search work is distributed across more workers, allowing the cost of searching more graphs to be amortized through parallelism. 

With a batch size of 32, Qdrant’s performance improves from 4 to 16 compute nodes, peaking at a mean QPS of 664 before tapering and eventually declining. With larger batch sizes of 256 and 2048, however, Qdrant continues scaling until 32 and 64 compute nodes, respectively. This suggests that batch size affects the balance between computation and communication overhead. All tested VDBs use a broadcast-gather query pattern where one worker broadcasts the query batch and then gathers/reduces the partial results. At small batch sizes, the per-query costs of communication, coordination, and reduction are amortized over relatively little computation, whereas larger batches better balance these costs. This is reflected in CPU utilization at 32 compute nodes: a batch size of 32 achieves consistently low CPU utilization of roughly 7\%, whereas a batch size of 2048 produces burstier but substantially higher utilization, ranging from 30\% to 70\%. These effects are likely dataset-dependent, and practitioners need to consider the best balance of computation and communication for their use case.

In contrast to Qdrant, Weaviate’s performance peaks at 4 compute nodes and then declines. Comparing the 4-node and 64-node configurations reveals significant workload imbalance that grows at scale. For the 4-node configuration, the merging worker exhibits CPU utilization of 28--50\%, while a non-merging worker remains at 10--15\%. This disparity suggests that non-merging workers have little useful work to perform, while the single merging worker increasingly becomes a bottleneck. This imbalance becomes more pronounced with 64 compute nodes: CPU utilization on the merging worker ranges from 5\% to 31\%, while a non-merging worker remains below 1\% utilization during query execution. The observed trend reflects a poor communication-to-computation ratio in which coordination and merging overhead dominates useful computation.

Despite our best efforts, we were unable to successfully upload the full Pes2o-VE dataset to Milvus for query evaluation. The developer-recommended pure Lustre deployment proved too susceptible to performance variability, leading to runtime failures caused by failed attempts to create WAL entries for inserts (observed with 8 and 16 compute nodes). We also tested the standard MinIO-based deployment, using Lustre as a storage backend because of storage requirements. With the standard insertion path and the Milvus bulk-upload utility, the MinIO configuration exhibited Lustre incompatibilities that resulted in data loss and runtime failures.

Together, the multinode query results highlight key limitations of the current VDB broadcast-gather query pattern. As the number of workers increases, the single aggregating worker can become a bottleneck, limiting scalability. This effect is most pronounced in Weaviate, whose throughput declines beyond 4 nodes. Qdrant scales more effectively, but its performance remains well below linear scaling and is sensitive to batch size, indicating that coordination and reduction overheads still constrain distributed query performance.

\begin{lessonlist}
\begin{itemize}[leftmargin=*, label=, itemsep=2pt]
    \item \textbf{Lesson 8:} The standard VDB broadcast-gather query pattern exhibits limited scaling and places an unequal load on the single aggregating worker, motivating the design of new distributed query strategies. 
\end{itemize}
\end{lessonlist}

\subsection{Evaluating Mixed Read/Write Workloads}
\label{sec:mixedWorkloads}
To provide a more complete picture of VDB performance, our final test evaluates a second workload pattern in which queries and insert requests are issued concurrently after an initial ingestion phase. We first insert 5 million vectors from Pes2o-VE and Yandex-T2I, allowing indexing to fully complete. This setup mimics a typical agentic workflow, where an agent begins with an initial knowledge base and later issues both queries and new inserts. We launch two independent clients: one sending queries and the other performing inserts (batch size = 32 for queries and inserts). 

Because of differences in insertion rates across systems, the total amount of data ingested during the experiment varies between VDBs. As a result, direct comparisons across VDBs are less meaningful if a given query was executed on a larger corpus. To accommodate this, we evaluate each VBD's relative performance degradation under load by identifying the point at which the final insert occurs and then comparing performance against a baseline in which the same workload (insert, full indexing, and query) is executed without concurrency. This approach allows us to isolate the impact of overlapping inserts and queries, rather than conflating results with differences in dataset size across VDBs. We repeat all experiments three times and report the mean and standard deviation.

As shown in \Cref{tab:mixedQueryInsert}, concurrent insertion and querying degrade performance across all VDBs. The impact, however, differs between throughput, latency, and tail latency. Across the tested datasets, Milvus experiences the smallest average throughput degradation, decreasing by 23.44\%, compared with 38.37\% for Weaviate and 50.57\% for Qdrant, while latency increases by 34.99\%, 102.74\%, and 66.52\% for Milvus, Qdrant, and Weaviate, respectively. In contrast, P99 latency increases much more sharply, rising by an average of 280.10\% on Pes2o-VE and 344.77\% on Yandex-T2I. This indicates that concurrent insertion has a disproportionate effect on the slowest queries, creating high-latency outliers. 

The significant rise in P99 latency reflects how each VDB handles newly inserted data during query execution. Milvus and Qdrant place new vectors in unindexed segments that must be searched with an exhaustive scan, increasing query latency. Weaviate takes a different approach: rather than using exhaustive search for newly inserted vectors, it directly updates the active HNSW index while queries are running, creating contention for shared CPU and memory resources. Furthermore, as is typical of mixed read-write workloads, concurrent insertion introduces lock contention across the tested VDBs: queries must be prevented from reading segment or index state while it is being modified. This synchronization cost is especially visible in tail latency: although most queries may execute without substantial delay, queries that arrive during an active mutation can block, causing P99 latency to rise sharply.

\begin{lessonlist}
\begin{itemize}[leftmargin=*, label=, itemsep=2pt]
    \item \textbf{Lesson 9:} Concurrent inserts affect query P99 latency more than mean latency or throughput, suggesting that latency-sensitive applications may need specialized data structures to preserve query SLAs during ingestion. 
\end{itemize}
\vspace{-6pt}
\end{lessonlist}

\section{Discussion and Conclusion}
Through large-scale deployment and evaluation of three VDBs on two HPC systems, several key insights emerge. We find that, compared with the cloud, HPC resources significantly improve indexing time and query throughput; however, they provide limited benefits for already low-latency queries (e.g., sub-ms). We observe that modern VDBs introduce a recall-cost trade-off by segmenting data across multiple indices. Between VDBs, we find that Milvus's indexing scales more effectively with core count than Qdrant, while Weaviate lags in indexing performance. Furthermore, our results show that, across VDBs, query performance frequently plateaus as cores increase and, in multinode settings, high communication costs limit multinode scaling. Beyond performance, we identify write amplification as a critical, underappreciated cost. Overall, we find that existing VDBs are frequently misaligned with HPC environments and are unable to fully harness HPC system resources.


Our findings indicate that researchers deploying VDBs should consider the influence of embedding geometry on potential HPC benefits, carefully configure their systems to avoid WAL bottlenecks during insertion, and explore different core counts to maximize query throughput. Additionally, our observations  point to future research directions for VDBs. To overcome the significant underutilization of network resources during insertion, future VDBs can explore ingestion APIs with multiple data paths and tunable consistency–performance trade-offs for large-scale scientific datasets. Moreover, given that query performance exhibits limited scaling with core counts, our work shows that new threading and distribution models are required to effectively utilize HPC resources.\footnote{ChatGPT \cite{chatgpt2026} was used to improve the grammar and phrasing of this work. }


\section*{Acknowledgment}
This material is based upon work supported by the U.S. Department of Energy (DOE), Office of Science, Office of Advanced Scientific Computing Research, including the TIDES project at Argonne National Laboratory, under Contract No. DE-AC02-06CH11357. This research also used resources of the Argonne Leadership Computing Facility, a DOE Office of Science user facility. Additionally, this material is based upon work supported by the National Science Foundation Graduate Research Fellowship Program under Grant No. 2137424. Any opinions, findings, and conclusions or recommendations expressed in this material are those of the author(s) and do not necessarily reflect the views of the National Science Foundation.

% ...
\small
\bibliographystyle{IEEEtran}
\bibliography{main}

\end{document}